\documentclass[final,3p,times,authoryear]{elsarticle}

\usepackage{amssymb}
\usepackage{amsmath}
\usepackage{graphicx}
\usepackage{xcolor}

\journal{Journal of Fluids and Structures}

\begin{document}

\begin{frontmatter}




\title{Dynamic mode decomposition based analysis of flow over a sinusoidally pitching airfoil}


\author{Karthik Menon\fnref{label1}}
\author{Rajat Mittal\corref{cor1}\fnref{label1}}
\cortext[cor1]{Corresponding author}
\ead{mittal@jhu.edu}

\address[label1]{Department of Mechanical Engineering, Johns Hopkins University, Baltimore, MD 21218, USA}

\begin{abstract}
Dynamic mode decomposition (DMD) has proven to be a valuable tool for the analysis of complex flow-fields but the application of this technique to flows with moving boundaries is not straightforward. This is due to the difficulty in accounting in the DMD formulation, for a body of non-zero thickness moving through the field of interest. This work presents a method for decomposing the flow on or near a moving boundary by a change of reference frame, followed by a correction to the computed modes that is determined by the frequency spectrum of the motion. The correction serves to recover the modes of the underlying flow dynamics, while removing the effect of change in reference frame. This method is applied to flow over sinusoidally pitching airfoils, and the DMD analysis is used to derive useful insights regarding flow-induced pitch oscillations of these airfoils. 
\end{abstract}


\end{frontmatter}


\section{Introduction}
\label{sec:intro}

Problems in fluid dynamics are often characterized by a range of time and length scales, which may be a result of the inherent non-linearity of the governing equations. In the particular case of fluid-structure interaction, the existence of natural scales associated with the structure, and its interaction with the flow scales, introduces additional complexity into the problem. The identification of such flow structures and their associated timescales, have considerably advanced our understanding of the fluid dynamics of such problems, particularly in the domain of fluid-structure interactions \citep{Williamson1996VortexWake,Ellington1996Leading-edgeFlight,Govardhan2000,Singh2005Vortex-inducedModes,Onoue2016VortexPlate,Eldredge2019,Menon2019FlowAirfoil}. 

The development and use of modal decomposition techniques to identify such flow features has gained increasing popularity in recent years, especially due to advances that have allowed the generation of large experimental and numerical data sets \cite{Taira2017ModalOverview}. One such technique, Dynamic Mode Decomposition (DMD), uses snapshots of the flow-field and allows the identification of spatial structures with characteristic frequencies and growth/decay rates associated with these structures \citep{Rowley2009SpectralFlows,Schmid2010DynamicData}. For nonlinear dynamical systems, such as most fluid-dynamics problems, the spatial modes extracted by DMD are the eigenvectors of the best-fit linear operator that approximates the dynamics of the field. Details of this analytical tool, its mathematical properties, as well as some simple applications of this method are presented in \cite{Rowley2009SpectralFlows}, \cite{Schmid2010DynamicData}, \cite{Schmid2011ApplicationData}, \cite{Schmid2011ApplicationsDecomposition} and \cite{Chen2012VariantsAnalyses}. 

DMD has been used in a wide variety of applications including the wake of a circular cylinder \citep{Bagheri2013Koopman-modeWake}, separated flow \citep{Hemati2017De-biasingDatasets}, jets \citep{Rowley2009SpectralFlows,Schmid2011ApplicationsDecomposition}, in the analysis of stall control \citep{Mohan2017AnalysisDecomposition} and the wake of a flapped airfoil \citep{Pan2011DynamicalFlow}. These studies, as well as the majority of others in the literature, involve the use of DMD in problems with stationary boundaries. In the context of fluid-structure interactions, there have been far fewer applications of DMD primarily due to the fact that this method, without modification, cannot account for a body of nonzero thickness moving through the field of interest. It must be noted that this is not a problem in the case of an infinitely thin body, as demonstrated for a flapping membrane by \cite{Goza2018ModalFlapping}. However, the motion of bodies of nonzero thickness introduces spurious structures in the computed DMD modes due to the time-varying position of the body in the field of interest. On account of this, such studies have focused exclusively on limited regions of the flow-field that are outside the overall envelope of the structural motion, such as the wake behind immersed bodies. One such demonstration of DMD on the wake of a flapping membrane was carried out by \cite{Schmid2010DynamicData}, and the decomposition of the wake behind a flexible cantilevered beam was performed by \cite{Cesur2014AnalysisDMD}. However, by limiting the region of interest in the flow-field to the wake, relevant flow structures close to the immersed body as well as within the shear layer are ignored. 

This deficiency of DMD was recognized by \cite{Mohan2016ModelAirfoil}, who performed a decomposition of the flow near the suction-side of a plunging airfoil using DMD and POD, by analyzing a region of the flow-field moving with the airfoil. While the simulations in their case were performed using a grid moving with the airfoil, which does not encounter the problem of having a body moving through the field of interest, it is non-trivial to use such an approach in experimental datasets and other methods where the locations of data-acquisition are fixed in the lab frame. In particular, one such method where this difficulty is apparent is in immersed boundary methods, which have become increasingly popular in flow simulations of fluid-structure interactions due to their ability to handle complex geometries and arbitrary motions in a robust manner \citep{Mittal2005}. Further, we see that the decomposition of flow-fields in the reference frame of the moving body are dominated by the rotational/translational velocities that are a result of the non-inertial nature of this reference frame. Hence, it is useful to be able to decouple these effects from the underlying fluid dynamics. 

Lastly, recent work in some canonical fluid-structure interaction problems has shown that the energy transfer between the moving body and the fluid, which depends on the phase difference between the motion and force on the immersed body, can be a very useful tool in understanding and predicting the flow-induced dynamics of the body \citep{Morse2009PredictionMotion,Bhat2013StallNumbers,Kumar2016Lock-inCylinder,Menon2019FlowAirfoil}. However, this energy transfer is a frame-dependent quantity, which is in fact zero in the frame of the moving body. Hence, establishing a connection between the computed modes in the two frames is useful in this context. Nevertheless, it is clear from the work of \cite{Mohan2016ModelAirfoil} that DMD is a promising tool in the analysis of fluid-structure interaction problems, and the various scales associated with these problems. Hence the development of a robust method to decompose the flow near a moving boundary of non-zero thickness could prove to have wide-ranging applications. 

Since the introduction of DMD, numerous improvements and variants of the method have been proposed to address issues associated with it \citep{kutz2016dynamic}. Optimized DMD and sparsity-promoting DMD, introduced by \cite{Chen2012VariantsAnalyses} and \cite{Jovanovic2014Sparsity-promotingDecomposition} respectively, have been developed as methods to represent the dynamics using the fewest possible modes. Efforts to perform DMD on large data sets have led to the development of more computationally efficient methods, such as those of \cite{Hemati2014DynamicDatasets} and \cite{Gueniat2015ASystems}. Ideas from compressed sensing have also been used in conjunction with DMD to make the method more tractable in large problems that are constrained by practical data-sampling rates \citep{Tu2014SpectralData,Brunton2015CompressedDecomposition}. Further, the effect of noise, such as experimental error or sensor noise, on the computed DMD modes as well as possible improvements have been addressed by \cite{Duke2012AnDecomposition}, \cite{Dawson2016CharacterizingDecomposition}, and \cite{Hemati2017De-biasingDatasets}. These advances have helped to address a variety of issues, and to widen the applicability of DMD for practical problems. However, as with the original formulation of DMD, these variants do not address the difficulty involved in accounting for moving boundaries in the flow-field. 

\begin{figure}
    \centering
    \includegraphics[scale=1.0]{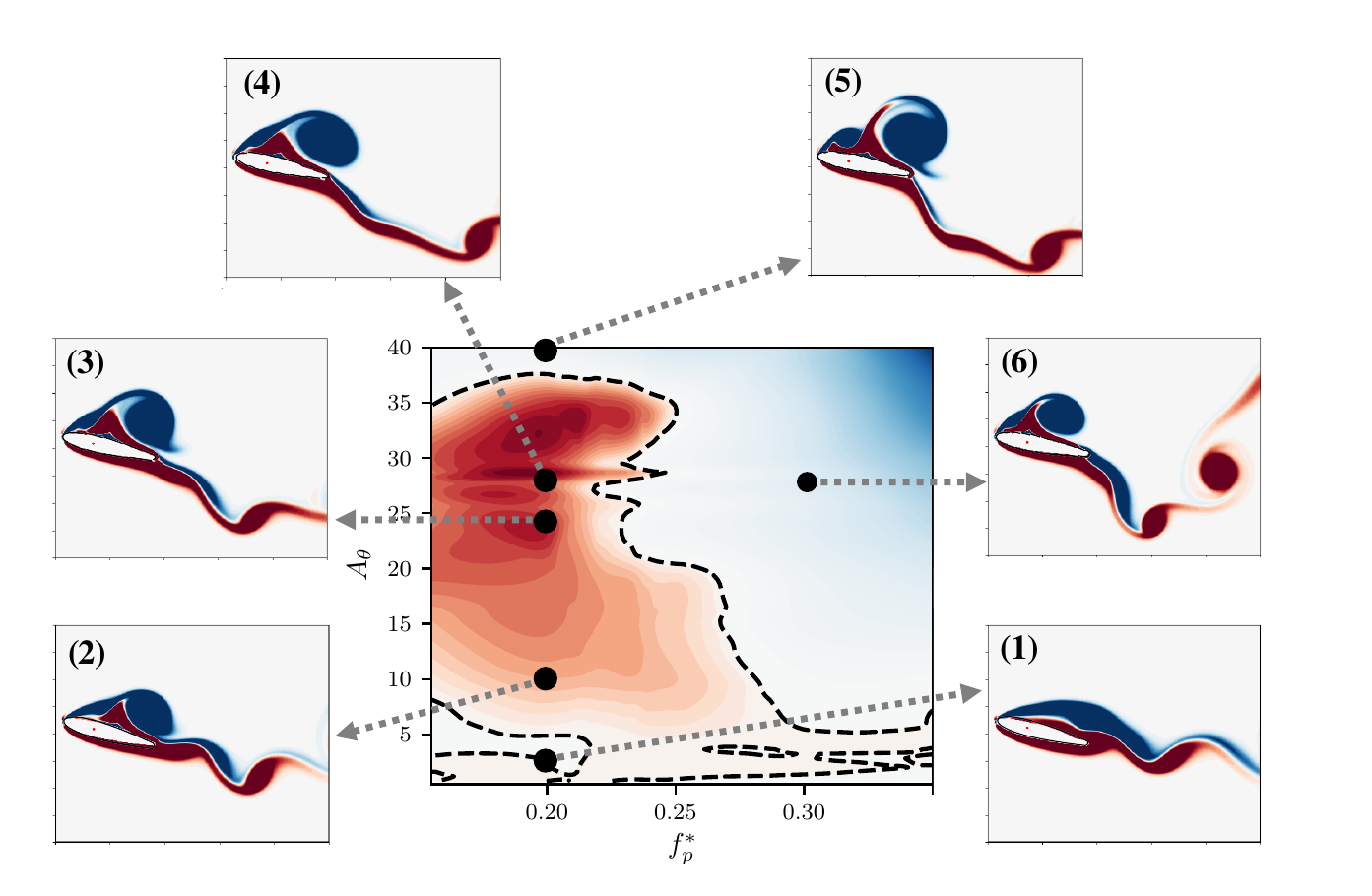}
    \caption{(a) Map of energy extraction ($C_E$) by a pitching airfoil from the surrounding flow, as a function of oscillation amplitude ($\Delta \theta$) and frequency ($f^*_p$). Reproduced from \cite{Menon2019FlowAirfoil}. Black circles denote the amplitude and frequency of cases discussed in section \ref{sec:airfoil} of this work. Vorticity snapshots at the time-instance corresponding to the mean position during pitch-down are shown for each of these cases.}
    \label{fig:energy_map}
\end{figure}{}
The particular moving-boundary problem of interest in this work is the flow around a sinusoidally pitching airfoil. As mentioned earlier in this section, our previous work \citep{Menon2019FlowAirfoil} has shown that the response of a freely-pitching airfoil, with a given structural frequency, is determined by the energy extracted by the airfoil from the surrounding flow at its instantaneous oscillation frequency and amplitude. The sign of the energy transfer determines whether the flow-induced oscillations grow or decay. This energy extraction for a pitching airfoil depends primarily on the phase angle between the oscillation and the pitching moment generated on the airfoil. We can show this phase dependence for a simple case of an airfoil performing sinusoidal pitching oscillations and experiencing sinusoidal pitching moment as follows. As in \cite{Menon2019FlowAirfoil}, the energy extracted by the airfoil from the surrounding flow over one oscillation cycle is given in terms of a energy coefficient $C_E$ as
\begin{equation}
  C_E = \int_{t}^{t+T} C_M \Omega dt
  \label{eq:energy_ext}
\end{equation}
where $\Omega$ is the angular velocity of the pitching airfoil, $T$ is the period of oscillation, and $C_M$ is the coefficient of moment on the airfoil. For this simple example, $\Omega = A_{\omega} \sin (2 \pi t/T)$ and $C_M = A_M \sin (2 \pi t/T +\phi)$. Plugging this kinematics and forcing into equation \ref{eq:energy_ext}, we see that the energy extraction is given by $C_E  = \left(1/2\right) A_{\omega} A_M T\cos\phi$. This shows that the sign of energy extraction, which governs the growth or decay of flow-induced oscillations, is determined by the phase difference, $\phi$ in this example, between the kinematics and forcing. 

\cite{Menon2019FlowAirfoil} also showed that by using prescribed oscillations at a range of frequencies and amplitudes, the computed energy transfer can be used to predict all branches of the free-vibration response within that range of frequency and amplitude. The bifurcation map of this system as a function of oscillation amplitude and frequency can then be created using the energy extraction, which was referred to as an "energy map". In figure \ref{fig:energy_map} we reproduce the energy map from the work of \cite{Menon2019FlowAirfoil} for an airfoil oscillating about a hinge location at $33\%$ of the chord. We see that there are regions of positive and negative energy extraction, separated by a curve of zero energy extraction (dashed line). The kinematic states with positive and negative energy extraction correspond to states with growing and decaying flow-induced oscillations respectively. The curve of zero energy extraction corresponds to equilibrium states for a structurally undamped flow-induced oscillator. Further, it must be pointed out that the energy map has a very complicated structure, with multiple regions of zero energy transfer for some oscillation frequencies, regions of high gradient in energy, and small "islands" of negative energy transfer. 

A key question that arises immediately is: what determines the topology of this energy map? We showed above that the energy transfer depends on timescales and phase differences between the flow and airfoil's kinematics. Hence we expect to be able to understand the topology of the energy map by analyzing the timescales associated with the dominant flow structures and their variation with airfoil kinematics. In figure \ref{fig:energy_map} we show snapshots of the instantaneous vorticity field for some cases on the energy map, selected from various regions of interest on the map. Apart from the growth of the leading-edge vortex (LEV) with oscillation amplitude and differences in the wake, it is evident that subtle differences in timing and phase between these cases are difficult to discern from the full flow-field. It is in this regard that the modal decomposition provided by a method such as DMD would be extremely useful. We are specifically interested in analyzing the phase and timing of flow structures in the decomposed flow with respect to the phase of the airfoil motion, in ways that are not feasible without accounting for the motion of the body.

In this work, we propose a method to decompose the flow around a moving boundary by transforming the analysis to a body-fitted frame. By exploiting the linearity of the formulation, we show rigorously how the modes computed in a moving coordinate system are related to those of the underlying flow dynamics in the lab-frame. We use this to apply a correction to the modes in order to decouple the effect of the moving-frame velocity from the flow dynamics, which helps to alleviate some of the issues outlined earlier in this section. Further we present a practical implementation of this method, by including the effect of changing coordinate systems, in the context of fluid-structure interactions using numerical simulation data from a sharp-interface immersed boundary method. It must be noted here that while the numerical method used in this work results in lab-frame data, which necessitates this change in coordinate systems on account of the moving body, it is possible for the raw flow-field data to correspond to the non-inertial frame \citep{Kim2006ImmersedBody}. The method presented here to distill the modes governing the underlying flow from the effects of the non-inertial frame have applications for these problems as well.

\section{Methods}
\label{sec:methods}

\subsection{Flow simulations}
\label{sec:num_meth}
The flow simulations in this study have been performed using the sharp-interface immersed boundary method of \cite{Mittal2008ABoundaries} and \cite{Seo2011AOscillations}. This method is particularly well-suited to fluid-structure interaction problems as it allows us the use of a simple non-conformal Cartesian grid to simulate a variety of different shapes and motions of the immersed body. Further, the ability to preserve the sharp-interface around the immersed boundary ensures very accurate computations of surface quantities. The incompressible Navier-Stokes equations are solved using a fractional-step method. Spatial derivatives are discretized using second-order central differences in space, and time-stepping is achieved using the second-order Adams-Bashforth method. The pressure Poisson equation is solved using a geometric multigrid method. This code has been extensively validated in previous studies \citep{Ghias2007,Mittal2008ABoundaries,Seo2011AOscillations}, where its ability to maintain local (near the immersed body) as well as global second-order accuracy has been demonstrated. Further, the accuracy of surface measurements has been established for a wide variety of stationary as well as moving boundary problems in these studies. 

\begin{figure}
    \centering
    \includegraphics[scale=1.0]{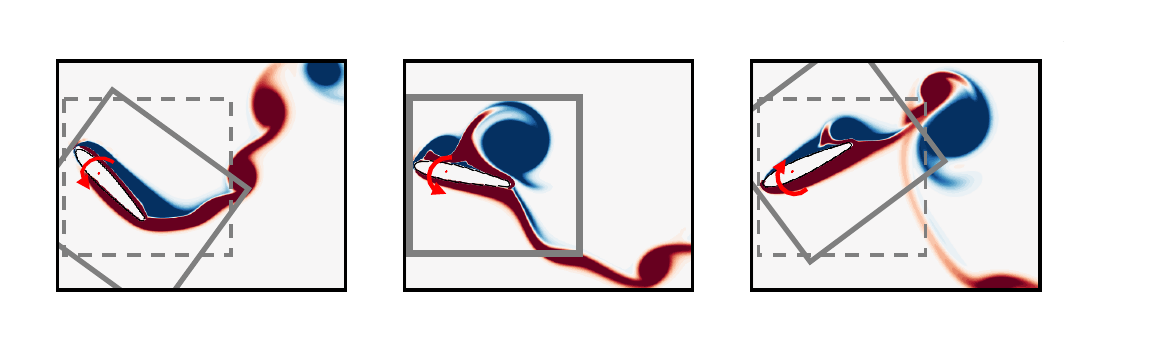}
    \caption{Schematic of the interrogation window for a sample case of a pitching airfoil, where the flow is visualized by contours of vorticity. Snapshots are recorded in the moving window (box with solid lines) and the lab-frame is represented using the window with dashed lines.}
    \label{fig:windows}
\end{figure}

The flow simulations reported in this work are performed assuming viscous, incompressible flow. The grid sizes used in the test cases reported, for cylinders and airfoils, are $320 \times 288$ and $384 \times 320$ cells respectively. The flow within a smaller region of this domain, which includes the immersed body at all times, is the focus of the mode decomposition described in this work. The size of this interrogation window is $n_x \times n_y = 120 \times 120$ and $250 \times 268$ grid points for the cylinder and airfoil cases respectively. This determines the size of the snapshots used in the computation of DMD modes. A schematic of the interrogation window is shown in figure \ref{fig:windows} using dashed lines. Flow measurements within this window are made on the Cartesian grid, and hence include points that are inside the immersed body (solid points) during some timesteps and outside (fluid points) during others. The DMD modes are computed using $250$ snapshots for most of the cases discussed. This corresponds to between $10$ and $15$ cycles of oscillation, depending on the oscillation frequency. We experimented with using as few as 1-2 cycles for computing the modes, however using too few snapshots was seen to yield modes with non-smooth structures. 

For the airfoil as well as cylinder cases demonstrated here, the motion of the moving body is prescribed, although the formulation presented is equally applicable to flow-induced motion. Details about the airfoil simulations can be found in \cite{Menon2019FlowAirfoil}. The Reynolds number is defined as $Re = \rho U_\infty L/\mu$, where $L$ is the reference length and the prescribed frequency of oscillation is reported in non-dimensional form as $f^*_p=f_pL/U_\infty$, where $f_p$ is the pitching frequency. All other frequencies are deonoted by $f^*=fL/U_\infty$. For the cases discussed here, the length-scale used is the chord-length ($L = C$) in the case of airfoils, and the diameter ($L=D$) in the case of cylinders. 
{}

\subsection{Dynamic mode decomposition}
\label{sec:dmd}
The DMD formulation used in this work, and summarized below, is based on the so-called  ``exact DMD'' of \cite{Tu2014OnApplications}. This is related to the Arnoldi-like formulation of \cite{Rowley2009SpectralFlows} and \cite{Schmid2010DynamicData}, but is known to be more numerically stable. This method takes snapshots of observables and computes the modes that govern the dynamics described by the sequence of observables. In this work, the observables correspond to flow-field measurements, separated by a time interval of $\Delta t$. They are represented as vectors $\boldsymbol{q_n} \in \mathbb{R}^{\beta \times N}$, where $N=n_x \times n_y$ is the number of grid points in each snapshot, $\beta$ is the number of flow variables included in the snapshot at every grid location, and $n$ is the index of the snapshot. The number of snapshots used in computing the decomposition is $N_t$, and these snapshots are organized as follows into two data matrices that are staggered in time: 
\begin{subequations}
    \begin{align}
        \boldsymbol{\mathrm{Q}} &= \big[ \; \boldsymbol{q}_0 \; | \; \boldsymbol{q}_1 \; | \; ... \; | \; \boldsymbol{q}_{N_t-1} \; \big] \label{eq:data_matrix}\\
        \boldsymbol{\mathrm{Q'}} &= \big[ \; \boldsymbol{q}_1 \; | \; \boldsymbol{q}_2 \; | \; ... \; | \; \boldsymbol{q}_{N_t} \; \big].
        \label{eq:data_matrix2}
    \end{align}
\end{subequations}

The idea is to find the best-fit linear operator for the dynamical system described by these observables, such that 
\begin{equation}
    \boldsymbol{q}_{n+1} = \boldsymbol{\mathrm{A}q}_n
    \label{eq:dynamical_sys}
\end{equation}{}
where $\boldsymbol{\mathrm{A}} \in \mathbb{R}^{N \times N}$ is a large matrix that governs this evolution. Note that this best-fit dynamical system that approximates the dynamics is a autonomous. Further, equation \ref{eq:dynamical_sys} can also be written in terms of the entire snapshot history as $\boldsymbol{\mathrm{Q'}}=\boldsymbol{\mathrm{AQ}}$, from which $\boldsymbol{\mathrm{A}}$ can, in theory, be directly computed as 
\begin{equation}
    \boldsymbol{\mathrm{A}}=\boldsymbol{\mathrm{Q'Q}}^{\dagger}
\end{equation}
where $\boldsymbol{\mathrm{Q}}^{\dagger}$ is the Moore-Penrose pseudo-inverse of $\boldsymbol{\mathrm{Q}}$. However, due to the large size of $\boldsymbol{\mathrm{A}}$, which makes it impractical to compute/store explicitly, a more computationally efficient method to proceed involves projecting the dynamics on the first $r$ principle components of the data matrix, $\boldsymbol{\mathrm{Q}}$. This is done via the Singular Value Decomposition of the data matrix, and by retaining the first $r$ principle components, $\boldsymbol{\mathrm{Q}} \approx \boldsymbol{\mathrm{U}}_r\boldsymbol{\mathrm{\Sigma}}_r \boldsymbol{\mathrm{V}}^*_r$, where the subscript $r$ refers to the number of components. We can now compute a more computationally tractable projection of $\boldsymbol{\mathrm{A}}$, given by 
\begin{equation}
    \boldsymbol{\mathrm{\Tilde{A}}} = \boldsymbol{\mathrm{U}}_r^*\boldsymbol{\mathrm{Q'}}\boldsymbol{\mathrm{V}}_r\boldsymbol{\mathrm{\Sigma}}_r^{-1}
\end{equation}

The eigendecomposition of $\boldsymbol{\mathrm{\Tilde{A}}}$ then yields eigenvalues $\lambda_k$ and eigenvectors $\boldsymbol{w}_k$ that satisfy $\boldsymbol{\mathrm{\Tilde{A}}}\boldsymbol{w}_k=\lambda_k\boldsymbol{w}_k$. \cite{Tu2014OnApplications} showed that the eigenvalues of $\boldsymbol{\mathrm{\Tilde{A}}}$ are equal to the eigenvalues of $\boldsymbol{\mathrm{A}}$, and the eigenvectors of $\boldsymbol{\mathrm{A}}$, denoted by $\boldsymbol{v}_k$, are related to those of $\boldsymbol{\mathrm{\Tilde{A}}}$ by
\begin{equation}
    \boldsymbol{v}_k=\frac{1}{\lambda_k}\boldsymbol{\mathrm{Q'}}\boldsymbol{\mathrm{V}}_r\boldsymbol{\mathrm{\Sigma}}_r^{-1}\boldsymbol{w}_k
\end{equation}

Having computed the eigenvalues and eigenvectors of $\boldsymbol{\mathrm{A}}$, we can express the DMD-approximated dynamics of the system as $\boldsymbol{q}_n = \boldsymbol{\mathrm{A}}\boldsymbol{q}_{n-1} = \boldsymbol{\mathrm{A}}^n\boldsymbol{q}_0$, which follows from equation \ref{eq:dynamical_sys}. Assuming each snapshot lies in the span of the eigenvectors of $\boldsymbol{\mathrm{A}}$, i.e., $\boldsymbol{q}_0 = \sum_k b_k \boldsymbol{v}_k$, we can connect this time-advancement of the system to the DMD modes as follows,
\begin{equation}
    \boldsymbol{q}_n = \boldsymbol{\mathrm{A}}^n\sum_k b_k \boldsymbol{v}_k = \sum_k \lambda^n_k b_k \boldsymbol{v}_k.
    \label{eq:discrete_time}
\end{equation}
Here, $b_k$ are the coefficients corresponding to the Galerkin projection of the DMD modes on the state of the system at the first snapshot, $\boldsymbol{q}_0$. Equation \ref{eq:discrete_time} hence describes the dynamics of the discrete map, given by equation \ref{eq:dynamical_sys}, to the DMD modes and eigenvalues. The time-advancement can also be written in a manner analogous to continuous-time systems as,
\begin{align}
    \boldsymbol{q}_n &= \sum_k b_k \boldsymbol{v}_k \exp{(\omega_kt_n)} \\
         &= \sum_k b_k \boldsymbol{v}_k \exp{\{\Re(\omega_k)t_n\}}\exp{\{i\Im(\omega_k)t_n\}}
    \label{eq:continuous_time}
\end{align}
where $\omega_k=\log(\lambda_k)/\Delta t$, and $\Re(\omega_k)$ and $\Im(\omega_k)$ are its real and imaginary parts. It is evident from equations \ref{eq:discrete_time} and \ref{eq:continuous_time} that the time-evolution of each DMD mode, $\boldsymbol{v}_k$, is determined by its corresponding eigenvalue $\lambda_k$, which is related to $\omega_k$ for each mode. Further, from equation \ref{eq:continuous_time} we see that the real and imaginary parts of $\omega_k$ determine the amplitude and frequency of each mode. In particular, the frequency of each mode can be written as $f_k = \Im(\omega_k)/2\pi$.

\subsection{Treatment of a moving body}
\label{sec:moving_bound}
We now describe a method to compute the DMD modes of the flow around a moving body based on the procedure outlined in section \ref{sec:dmd}, by using snapshots that include the time-varying position of the immersed body. Further, for the reasons outlined in section \ref{sec:intro}, we are specifically interested in the lab-frame DMD modes. The interrogation window of interest is shown using dashed lines in figure \ref{fig:windows}, and described in section \ref{sec:num_meth}. For simplicity, we first discuss the specific case of a body performing periodic rotation/pitch oscillations at one frequency. Further, we assume that the observables being used for the decomposition are the $X$- and $Y$- components of velocity. Generalizations of this method are discussed subsequently. 

\begin{figure}
    \centering
    \includegraphics[scale=1.3]{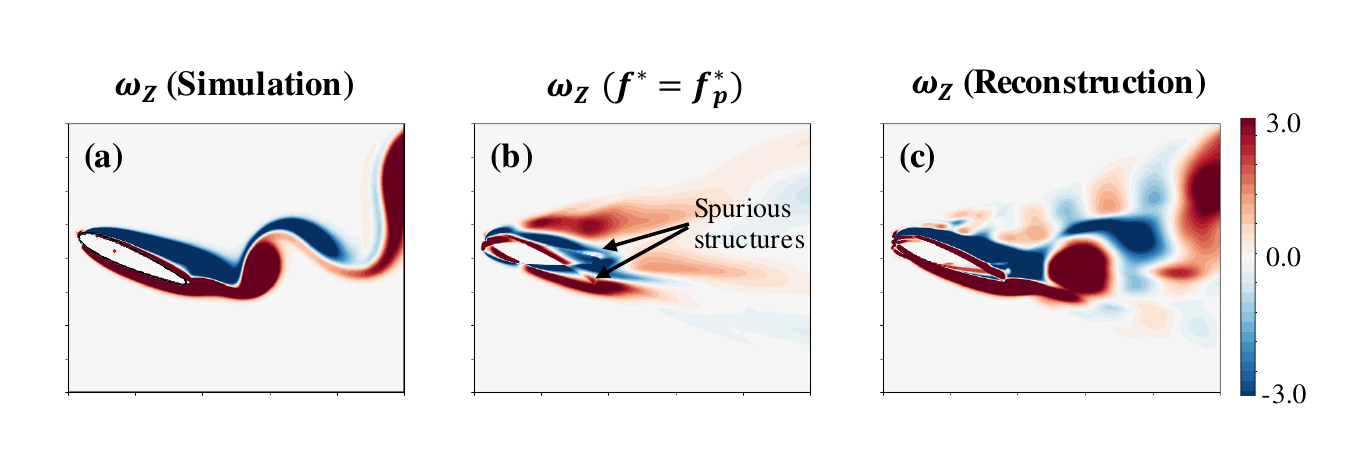}
    \caption{Application of DMD without the moving body formulation to the case of a sinusoidally pitching airfoil at $Re=1000$. The airfoil is oscillating about $\theta=15^{\circ}$ with amplitude and frequency of oscillation given by $\Delta \theta=10^{\circ}$ and $f^*_p=0.20$ respectively. All snapshots are at the time-instance corresponding to the the maximum pitch-up position. (a) Snapshot of $Z$-vorticity contours; (b) $Z$-vorticity contours for a computed DMD mode at the oscillation frequency, $f^*=f^*_p$. Spurious structures consisting of signatures of the airfoil surface at different pitch positions are highlighted; (c) Reconstruction of $Z$-vorticity field using 5 DMD modes.}
    \label{fig:direct_dmd}
\end{figure}

We begin with a brief demonstration of the direct application of the procedure described in section \ref{sec:dmd} on problems involving moving boundaries. As a sample case, we perform the decomposition on the flow around a sinusoidally pitching airfoil, oscillating about $\theta_0=15^{\circ}$ with amplitude and frequency of oscillation given by $\Delta \theta=10^{\circ}$ and $f^*_p=0.20$ respectively. Hence the instantaneous geometric angle of attack is given by $\theta = 15^{\circ} + 10^{\circ}\sin(2\pi \cdot 0.20 \cdot t)$. Figure \ref{fig:direct_dmd}(a) shows a snapshot of the flow for this case, coloured by contours of $Z$-vorticity, at a time-instance corresponding to the maximum pitch-up angle during the oscillation cycle. As mentioned above, we are interested in decomposing the lab-frame velocity field, given by $\boldsymbol{u}(\boldsymbol{x},t) = (u_x,u_y)$, where position-vectors are denoted by $\boldsymbol{x}$. The observables are vectors of the form, 
\begin{equation}
    \boldsymbol{q}_n \in \mathbb{R}^{2 \times N} : \boldsymbol{q}_n = [u_x(\boldsymbol{x}_1,t_n), ..., u_x(\boldsymbol{x}_{N},t_n),u_y(\boldsymbol{x}_1,t_n), ..., u_y(\boldsymbol{x}_{N},t_n)]^T.
    \label{eq:lab_observables}
\end{equation}
In figure \ref{fig:direct_dmd}(b) we plot the $Z$-vorticity of the computed DMD mode, corresponding to the oscillation frequency $f^*=f^*_p=0.20$, at the same time-instance as the flow snapshot in figure \ref{fig:direct_dmd}(a). We see the presence spurious structures within the region of the window that includes the time-varying position of the body. Specifically, a signature of the airfoil surface in its maximum pitch-up and pitch-down positions appears in the computed mode shape. These structures are highlighted using arrows in figure \ref{fig:direct_dmd}(b). Furthermore, the actual instantaneous shape of the airfoil is not retained in the computed mode and therefore the vortex structures that appear inside the actual airfoil shape are also spurious.  Further, in figure \ref{fig:direct_dmd}(c) we show the reconstruction of the flow using 5 DMD modes, at the same time-instance as that discussed above. We see that the shape of the airfoil is inaccurately reconstructed, which in turn causes inaccuracies in the reconstruction of the shear layer on the surface of the airfoil. This is an effect that is expected, and was alluded to by \cite{Goza2018ModalFlapping}. The inaccurate reconstruction around the moving boundary stems from the fact that the points within the overall envelope of the body's motion switch between being solid points and fluid points as the body moves. Hence the presence of the moving body appears as a diffused structure, as would be expected in the sense of an overall ``average'' position of the body. In mathematical terms, recall from equation \ref{eq:dynamical_sys} that the dynamical system being approximated by this method is necessarily autonomous. However, due to the fact that the fluid domain corresponds to different positions in each snapshot, which leads to time-varying discontinuous boundaries in $q_n$, the snapshots given by equation \ref{eq:lab_observables} do not satisfy this condition. 
{}

In order to circumvent the issue associated with the motion of the body within the lab-frame, we first make a coordinate transformation to a body-fixed reference frame. Snapshots of the flow in this moving-frame are recorded by using an auxiliary interrogation window that moves with the body, where the immersed body appears stationary. A schematic of this moving window is shown in figure \ref{fig:windows} using solid lines. Additionally, this non-inertial reference frame, in general, includes rotational and/or translational velocities that are related to the motion of the body with respect to the lab frame. Due to the fact that the body appears stationary in this moving frame, a first step is the straightforward application of the procedure outlined in section \ref{sec:dmd} to compute the DMD modes of the flow around the body in the body-fixed frame. We will then show how these modes are related to the decomposition in the lab-frame.

In the present work, this coordinate transformation is done by translating and/or rotating an auxiliary Cartesian grid, of size $N'$ grid points, with the body. For reasons that will become apparent later in this section, this moving window is necessarily larger in size than the lab-frame interrogation window as it is required to include the spatial extent of the lab-frame window at all times. We shall refer to position vectors in this moving-frame by $\boldsymbol{\xi}$, and the flow-field velocity is denoted by $\boldsymbol{u'}(\boldsymbol{\xi},t)$. We calculate the velocity field in the moving frame, at $\boldsymbol{\xi}$-coordinates, from the simulation data at locations in the lab-frame by using a bilinear interpolation scheme, and taking into consideration the relative velocity between the two frames.

We can now perform the decomposition of the velocity-field in the moving frame. Observables in this frame, $\boldsymbol{q'}_n \in \mathbb{R}^{2\times N'}$ take the same form as in equation \ref{eq:lab_observables}, but are larger and consist of velocities $\boldsymbol{u'}(\boldsymbol{\xi},t_n) = (u'_x,u'_y)$. The decomposition yields moving-frame DMD eigenvalues and modes and we can write the observables in this frame, $\boldsymbol{q'}_n$, in terms of these eigenvalues and modes as was done in equation \ref{eq:continuous_time}. This is shown below,
\begin{align}
    \boldsymbol{q'}_n &= \sum_k b_k \boldsymbol{v'}_k \exp{(\omega_kt_n)} \\
    &= \sum_k b_k \boldsymbol{v'}_k \exp{\{\Re(\omega_k)t_n\}}\exp{\{i\Im(\omega_k)t_n\}}
\end{align}
where it is implicitly assumed that $\boldsymbol{q'}_n = \boldsymbol{q'}(\boldsymbol{\xi},t_n)$ and $\boldsymbol{v'}_k = \boldsymbol{v'}_k(\boldsymbol{\xi})$, i.e., they refer to quantities in the moving frame. Here, $\omega_k$ and $\boldsymbol{v'}_k$ are the computed DMD eigenvalues and modes in this frame, and $b_k$ is the projection of these modes on the first snapshot, $\boldsymbol{q'}_0$. The growth and decay of these modes is governed by $\Re(\omega_k)$, and it is seen that for the decomposition of periodic flows in a stationary state, the dynamically relevant modes have non-decaying amplitudes. Hence it is reasonable to assume that $\exp{[\Re(\omega_k)t]} \approx 1$, which then yields the following form for the decomposition:
\begin{equation}
    \boldsymbol{q'}_n(\boldsymbol{\xi})= \sum_k b_k \boldsymbol{v'}_k \exp{\{i\Im(\omega_k)t_n\}}
    \label{eq:moving_dmd_moving}
\end{equation}

We now have the decomposition of the flow-field in the moving-frame, $\boldsymbol{u'}(\boldsymbol{\xi},t)$. However, as mentioned previously we are interested in the modes of the lab-frame flow-field, $\boldsymbol{u}(\boldsymbol{x},t)$, i.e., the modes of the observables $\boldsymbol{q}(\boldsymbol{x},t)$.  We can write $\boldsymbol{u}(\boldsymbol{x},t)$ in terms of the moving-frame velocity-field using the relative motion between the moving-frame and lab-frame coordinates, which we refer to as $\boldsymbol{\Tilde{r}}$. The relationship in lab-frame coordinates is as follows:
\begin{equation}
    \boldsymbol{u}(\boldsymbol{x},t) = \boldsymbol{u'}(\boldsymbol{x},t)+\boldsymbol{\Tilde{r}}(\boldsymbol{x},t).
    \label{eq:lab_moving_relation_vel}
\end{equation} 
For the sake of this formulation, it is clearer to express this relationship in terms of the observables used in the computation of DMD modes. In this form the relative-velocity field, $\boldsymbol{\Tilde{r}}(\boldsymbol{x},t_n)$, is written as a vector $\boldsymbol{r}_n(\boldsymbol{x})$, such that its size and arrangement of grid points is determined by that of $\boldsymbol{q}_n(\boldsymbol{x})$. 
\begin{equation}
    \boldsymbol{q}_n(\boldsymbol{x}) = \boldsymbol{q'}_n(\boldsymbol{x})+\boldsymbol{r}_n(\boldsymbol{x}).
    \label{eq:lab_moving_relation}
\end{equation}
As a first step in recovering the decomposition of the lab-frame velocity field, it must be noted that $\boldsymbol{q'}_n(\boldsymbol{x})$, i.e., the first term in the right-hand side of equation \ref{eq:lab_moving_relation}, only differs from $\boldsymbol{q'}_n(\boldsymbol{\xi})$ in the locations at which measurements are made. Since we have computed the decomposition of $\boldsymbol{q'}_n(\boldsymbol{\xi})$, in equation \ref{eq:moving_dmd_moving}, we can use a bilinear interpolation from the points in the moving-frame interrogation window to compute the modes of $\boldsymbol{q'}$ in the lab-frame coordinates. We must point out that this interpolation is the reason for the moving-frame interrogation window to necessarily include the spatial extent of the lab-frame window at all times, as was noted earlier. We refer to these interpolated modes as $\boldsymbol{v}_k(\boldsymbol{x})$. Hence the decomposition of $\boldsymbol{q}'$ can now be written in the lab-frame as:
\begin{equation}
    \boldsymbol{q'}_n(\boldsymbol{x})= \sum_i b_i \boldsymbol{v}_i \exp{[i\Im(\omega_i)t_n]}
    \label{eq:moving_dmd_lab}
\end{equation}

The second term in equation \ref{eq:lab_moving_relation} is a vector that, in general, includes the rotational and/or translational velocities of the points in the lab-frame interrogation window, as seen from the moving-frame. For the specific case of pure rotation, this velocity is $\boldsymbol{\Tilde{r}}(\boldsymbol{x},t) = \boldsymbol{\Omega}(t) \times \boldsymbol{x} = (-\Omega y,\Omega x)$. This can be written in the form of the observables (see equation \ref{eq:lab_observables}) as a vector $\boldsymbol{r}_n = \boldsymbol{p}\Omega(t_n)$, such that $\boldsymbol{p}$ is a vector containing the $X$ and $Y$-coordinates of all points in the interrogation window in the form:  
\begin{equation}
    \boldsymbol{p} \in \mathbb{R}^{2 \times N} : \boldsymbol{p} = [-y(\boldsymbol{x}_1), ..., -y(\boldsymbol{x}_{N}),x(\boldsymbol{x}_1), ..., x(\boldsymbol{x}_{N})]^T.
    \label{eq:coord_vector}
\end{equation}
Further, assuming a single frequency of oscillation for simplicity, given by $\omega_0$, the angular velocity can be written as $\Omega(t) = A_{\omega} \cos{(\omega_0t + \phi)}$. Using this,we can express $\boldsymbol{r}_n$ in the following form:
\begin{equation}
    \boldsymbol{r}_n = \boldsymbol{p}\Omega(t_n) = \boldsymbol{p} \big[A_{\omega} \cos{\omega_0t_n + \phi} \big]
    \label{eq:rot_simple}
\end{equation}
\begin{equation}
    \boldsymbol{r}_n = \boldsymbol{p} A_\omega \Bigg[\frac{\exp{\{i(\omega_0t_n+\phi)\}} + \exp{\{-i(\omega_0t_n+\phi)\}}}{2} \Bigg]
    \label{eq:rot_exp}
\end{equation}
From equation \ref{eq:rot_exp} we see that the relative-velocity vector can be written in terms of oscillating complex exponentials, much like the form for the the decomposed velocity, in equation \ref{eq:moving_dmd_lab}. This suggests we can use the linearity of the decomposition to recover the modes of the lab-frame velocity from the computed modes in the moving-frame. Plugging equations \ref{eq:moving_dmd_lab} and \ref{eq:rot_exp} into equation \ref{eq:lab_moving_relation}, we get,
\begin{equation}
    \boldsymbol{q}_n(\boldsymbol{x})= \sum_k b_k \boldsymbol{v}_k(\boldsymbol{x}) \exp{\{i\Im(\omega_k)t_n\}} + \boldsymbol{p} A_\omega \Bigg[\frac{\exp{\{i(\omega_0t_n+\phi)\}} + \exp{\{-i(\omega_0t_n+\phi)\}}}{2} \Bigg]
    \label{eq:dmd_rot_combined}
\end{equation}
The first term in equation \ref{eq:dmd_rot_combined} is a summation over all modes computed in the decomposition, which includes the modes computed at the oscillation frequency and its complex conjugate, which we denote as $v_{\omega_0}$ and $v_{-\omega_0}$ respectively. Since this is a linear superposition, we can take these modes out of the summation, and combine them with the second term in equation \ref{eq:dmd_rot_combined} where the complex exponential has the same form as that multiplying the DMD modes. Doing this, we can rewrite equation \ref{eq:dmd_rot_combined} as shown below:
\begin{align}
    \boldsymbol{q}_n(\boldsymbol{x}) = \sum_{\omega_k \neq \pm \omega_0} &b_k \boldsymbol{v}_k(\boldsymbol{x}) \exp{[i\Im(\omega_k)t_n]} \nonumber\\ 
    + \Bigg[ &b_{\omega_0}\boldsymbol{v}_{\omega_0} + \frac{A_{\omega}}{2}\boldsymbol{p} \exp{(i\phi)} \Bigg] \exp{\{i\Im(\omega_0)t_n\}} \nonumber\\ 
    + \Bigg[ &b_{-\omega_0}\boldsymbol{v}_{-\omega_0} + \frac{A_{\omega}}{2}\boldsymbol{p} \exp{(-i\phi)} \Bigg] \exp{\{-i\Im(\omega_0)t_n\}}
    \label{eq:dmd_moving_final}
\end{align}
Equation \ref{eq:dmd_moving_final} now gives us a form for the decomposition of the lab-frame velocity in terms of the modes computed in the moving frame. We see that the correction to the modes at the oscillation frequency, which come from the rotation term, have a non-trivial form that depends on the magnitude of angular velocity, the vector of $y$-coordinates, and the phase of this rotation, $\phi$. Another interesting observation from equation \ref{eq:dmd_moving_final} is that this change in reference frame only modifies the modes at the oscillation frequency. We will show in a practical example that this is true. 

While we have demonstrated this formulation for a body rotating at one frequency for simplicity, this method can be generalized for bodies performing more complicated combinations of periodic translation and rotation in a straightforward manner. This is done by including the terms for the translational velocity in equation \ref{eq:lab_moving_relation}, and writing those terms as complex exponentials as in equation \ref{eq:rot_exp}. Further, quasi-periodic motions can be treated in a similar manner by including all modes of oscillation in equation \ref{eq:rot_simple}. This will result in corrections to the DMD modes corresponding to all frequencies in the immersed body's oscillation frequency spectrum.

\section{Applications}
\label{sec:applications}

\subsection{Rotating cylinder}
\label{sec:cylinder}
As a first application of the methods presented in section \ref{sec:methods}, we analyze the decomposition of the flow around a circular cylinder performing periodic pitch (i.e. rotational) oscillations. This test case is particularly valuable in this study as it is one of few canonical ``moving boundary'' problems (although it can be implemented using a stationary surface and oscillating boundary conditions) which can be analyzed using the standard DMD formulation of section \ref{sec:dmd}, as the boundary does not sweep through the fluid domain.

In performing the two variants of DMD on this flow, we use an interrogation window of size $2D \times 2D$, where $D$ is the diameter of the cylinder, and the window is centered on the cylinder. The cylinder is forced to perform pitch oscillations of amplitude $\Delta \theta = 10^{\circ}$ and frequency $f^*_p=0.25$. The Reynolds number used here is $Re=100$, and the frequency of lift force oscillations on the cylinder is measured to be $f^* \approx 0.17$. This gives rise to two important timescales in the flow, i.e., dynamics governed by the oscillation frequency and that of the forcing frequency. We analyze the computed DMD modes at these frequencies, and compare results from the two methods. Comparisons are made at four phases during one oscillation cycle - the mean position during pitch-up, the peak pitch-up position, the mean position during pitch-down, and the peak pitch-up position.

\begin{figure}
    \centering
    \includegraphics[scale=1.0]{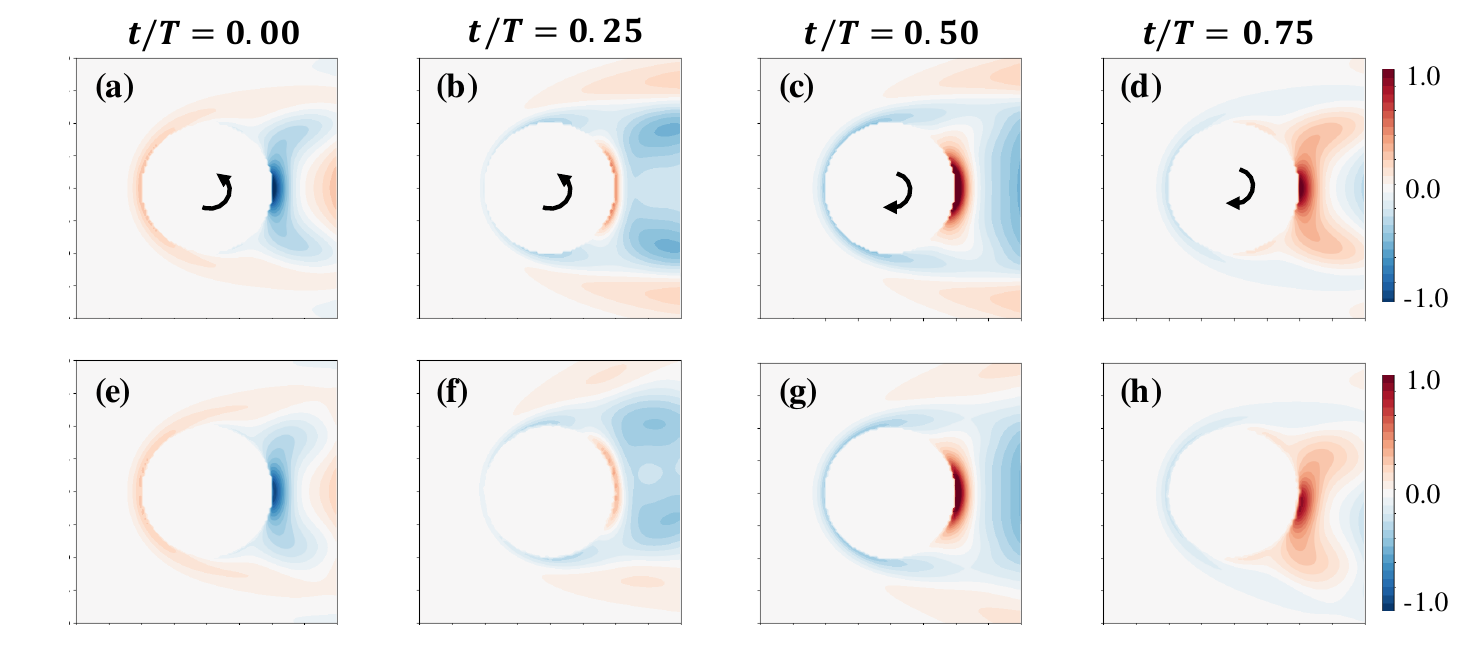}
    \caption{$Z$-vorticity contours of the DMD mode corresponding to $f^*=0.17$, for the flow around a circular cylinder at $Re=100$, $f^*_p=0.25$ and $\Delta \theta = 10^{\circ}$. The mode is plotted at four different phases of one oscillation cycle. The text above each panel shows the corresponding time-instance, where $T=1/f^*_p=1/0.25$. Arrows show the direction of rotation of the cylinder; (a)-(d) Modes computed using standard DMD; (e)-(h) Modes computed using moving-body formulation.}
    \label{fig:cyl_17}
\end{figure}{}
Figure \ref{fig:cyl_17} shows snapshots of the DMD mode at $f^*=0.17$ during these phases of the oscillation cycle. The top panel, figures \ref{fig:cyl_17}(a)-(d), show the modes computed using the standard DMD, and figures \ref{fig:cyl_17}(e)-(h) show modes computed using the moving-body formulation. We see that the modes have the same topology over time in the two cases. The only difference is the rotation of the mode in the case of the moving body formulation, which is a result of the moving window. This serves as a confirmation of one finding from equation \ref{eq:dmd_moving_final}, that the modes that do not correspond to the oscillation frequency were found to be unmodified. Further, we observe that the vortex structures at this frequency (which is the frequency of lift-forcing) are relatively stronger in the wake than in the shear layer. This suggests that the frequency of vortex shedding in the wake is closely related to the frequency of lift-forcing on the cylinder - a fact that has been widely used in studies of vortex-induced vibration to estimate the shedding frequency and/or identify lock-in \citep{Kumar2016Lock-inCylinder}. 

\begin{figure}
    \centering
    \includegraphics[scale=1.0]{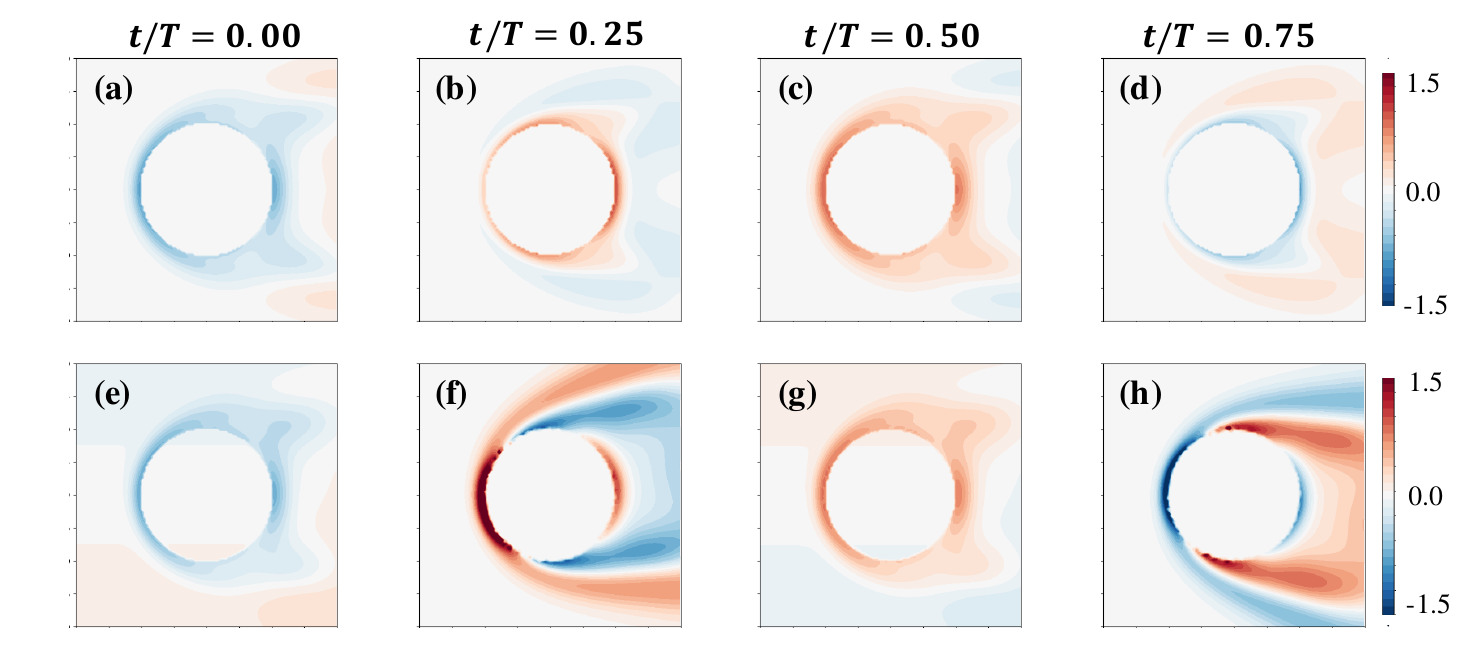}
    \caption{$Z$-vorticity contours of the DMD mode corresponding to $f^*=0.25$, for flow around a circular cylinder with the same parameters as in figure \ref{fig:cyl_17}. The mode is plotted during different phases of one oscillation cycle. Time-instances and direction of rotation are as in figure \ref{fig:cyl_17}; (a)-(d) Modes computed using standard DMD; (e)-(h) Modes computed using moving-body formulation.}
    \label{fig:cyl_25}
\end{figure}{}
In figure \ref{fig:cyl_25} we plot the DMD mode at $f^*=0.25$, at the same instances in time as in the preceding discussion. Figures \ref{fig:cyl_25}(a)-(d) correspond to the standard DMD, and figures \ref{fig:cyl_25}(e)-(h) correspond to the moving body formulation. We see from figures \ref{fig:cyl_25}(a) and \ref{fig:cyl_25}(e), as well as figures \ref{fig:cyl_25}(c) and \ref{fig:cyl_25}(g), that the modes computed by the moving-window and standard formulation share the same structure at the mean position in the oscillation cycle. This is due to the fact that the rotational effects in the flow-field are taken into account in the formulation for the moving window. Additionally, the interrogation windows overlap at this mean position, which removes any spatial differences between the flow seen from the two frames. At the other two time instances shown, which are at the peak pitch-up and pitch-down positions, the moving and stationary windows are at their maximum rotation angle with respect to each other. Hence spatial differences in the decomposed flow seen from the two frames are most evident at these time instances, which gives rise to qualitatively different structures (the effects of the rotational velocity field are minimum at these time-instances). Both frames show much larger structures around the cylinder surface at this frequency, which are also convected into the wake. The presence of large structures around the shear layer at this frequency is expected, as this corresponds to the frequency of oscillation of the cylinder. Hence we see from this example that even the decomposition of flows in relatively simple fluid-structure interaction problems provides valuable insight into flow modes associated with the different timescales.

\subsection{Sinusoidally Pitching Airfoil}

\subsubsection{Flow reconstruction}
\label{sec:reconst}
An important application of modal-decomposition techniques is the development of low-order models of the dynamics. This is based on the idea that most of the physically relevant dynamics are governed by a small number of modes. In this section, we demonstrate the reconstruction of the flow-field around a pitching airfoil via the DMD modes computed by using the moving-body formulation, which serves as a simple verification of the decomposition method described in this work. The specific case discussed here is that of an airfoil performing periodic pitching oscillations at frequency $f^*_p=0.20$ and amplitude $\Delta \theta = 10^{\circ}$. The Reynolds number is $Re = 1000$, and the interrogation window is of size $2.5C \times 2C$. We focus on the reconstruction of the vorticity field.
\begin{figure}
    \centering
    \includegraphics[scale=1.0]{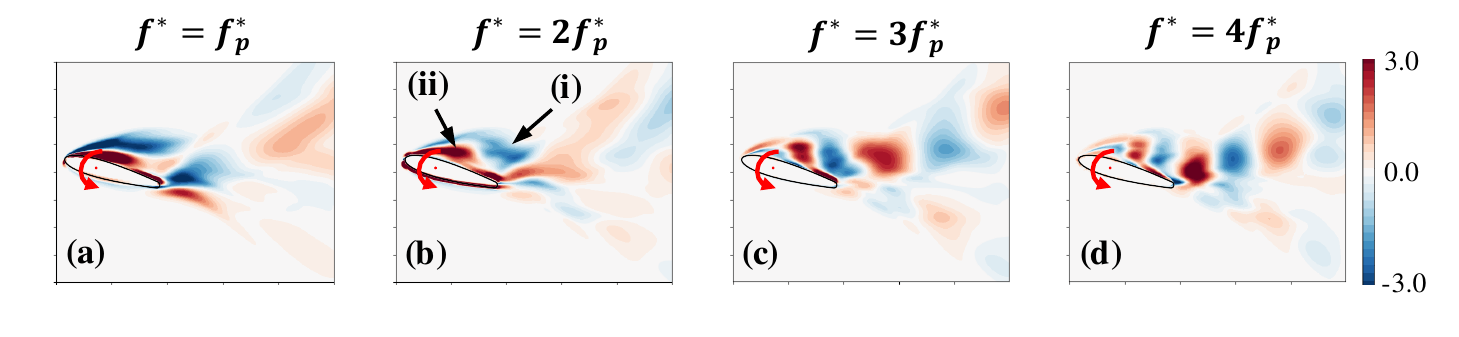}
    \caption{DMD modes of the flow around a pitching airfoil at $Re=1000$, pitching with $f^*_p=0.20$ and $\Delta \theta = 10^{\circ}$. The modes are coloured by contours of $Z$-vorticity, and the time-instance corresponds to the mean position during the pitch-down phase of oscillation. Modes at the oscillation frequency and its harmonics are shown. (a) $f^* = f^*_p$; (b) $f^* = 2f^*_p$; (c) $f^* = 3f^*_p$; (d) $f^* = 4f^*_p$. Note that this is the same case as that shown in figure \ref{fig:direct_dmd}.}
    \label{fig:airfoil_reconst_modes}
\end{figure}{}

We first analyze the flow physics captured by the dynamically important modes that are used in the subsequent reconstruction. Here we assess the relative importance of each mode by the magnitude of its Galerkin projection on the full flow-field. In figure \ref{fig:airfoil_reconst_modes} we show four leading modes, which are plotted at the time-instance corresponding to the mean position during the pitch-down motion. Note that, the moving boundary DMD formulation does not generate unphysical flow information near the airfoil surface as it did for the original formulation is figure \ref{fig:direct_dmd}. Figure \ref{fig:airfoil_reconst_modes}(a) shows the mode at the oscillation frequency, $f^*=f^*_p=0.20$, where the dominant structure is the separated shear layer over the suction surface of the airfoil and over the trailing edge. These shear layers are hence the driving mechanism at the fundamental frequency, and give rise to finer-scale vortex-structures at the higher harmonics. The first harmonic of this mode, $f^*=2f^*_p$ shown in figure \ref{fig:airfoil_reconst_modes}(b), captures the rolled-up vortex structures convecting over the surface of the airfoil (denoted by (i); negative vorticity) as well as the separating shear layer (denoted by (ii); positive vorticity) induced by this vortex. We see evidence of a leading-edge vortex (LEV) that is shed during the pitch-down motion at this frequency. The higher harmonics, $f^*=3f^*_p$ and $f^*=4f^*_p$, contain high wavenumber structures and are shown in figures \ref{fig:airfoil_reconst_modes}(c) and \ref{fig:airfoil_reconst_modes}(d). These structures correspond to vortices that form vortex-pairs and subsequently accelerate into the wake as a result of the jet induced between the vortices in each pair. It is interesting to note that these higher-harmonic modes, which are stronger in the wake, are approximately at the vortex shedding frequency of bluff-body wakes when using the projected area as the length scale ($f^* \approx 0.16$). 

\begin{figure}
    \centering
    \includegraphics[scale=1.15]{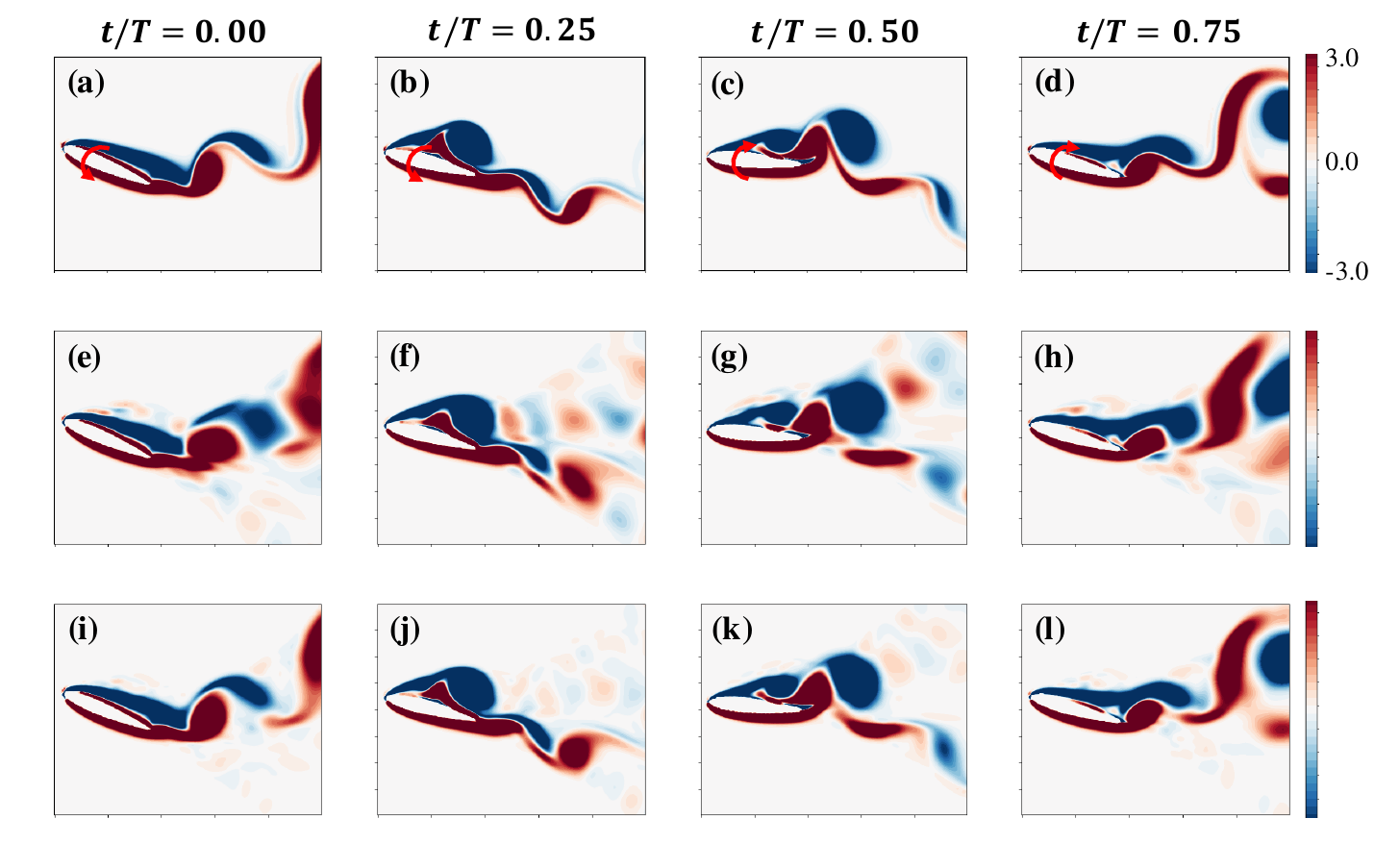}
    \caption{Contours of the $Z$-vorticity field around a pitching airfoil from the flow simulation data and from the reconstruction of the flow using DMD modes. The flow and kinematic parameters are $Re=1000$, $f^*_p=0.20$ and $\Delta \theta = 10^{\circ}$. Snapshots are shown at four phases over one oscillation cycle (time instance is specified at the top of each column). (a)-(d) Simulation data; (e)-(h) Reconstruction using five modes. This can be compared with the reconstruction in figure \ref{fig:direct_dmd}(c); (i)-(l) Reconstruction using ten modes. }
    \label{fig:reconst}
\end{figure}{}
We can now use just a few of the computed modes, i.e. the leading modes discussed above, to reconstruct the full flow-field. Figures \ref{fig:reconst}(a)-(d) show the vorticity field around the airfoil from the flow simulation data, as it goes through four phases during one oscillation cycle - the start of the pitch-down motion, the mean position during pitch-down, the start of pitch-up motion, and the mean position during pitch-up, respectively. The primary flow structures we see are the shedding of a leading-edge vortex during pitch-down, followed by the shedding of another smaller vortex later in the pitch-down motion. These vortices entrain the shear layer on the suction side and ``jet'' into the wake as vortex pairs. In figures \ref{fig:reconst}(e)-(h) we show the reconstruction of the flow at the same time-instances as in figures \ref{fig:reconst}(a)-(d), using just five DMD modes plus the mean (which corresponds to the $f^*=0$ mode). We see that these modes are able to capture the significant features of the vortex dynamics, i.e., the shedding and convection of the two vortices during the pitch-down motion. We see that there are also smaller-scale structures that are present in the reconstruction, but absent in the simulation data. These structures are associated with higher harmonic modes that are not included in the reconstruction, and are expected to have minimal dynamical significance in grid-converged simulations. Although not shown here, it must be noted that this reconstruction, when applied to the velocity field, shows smoother fields. 

We compare this 5-mode reconstruction with a reconstruction using ten DMD modes, plus the mean, in figures \ref{fig:reconst}(i)-(l). It is interesting to note that doubling the number of modes in the reconstruction does not significantly improve the reconstruction, especially very close to the airfoil, which confirms that the 5-mode reconstruction captures the bulk of the dynamics. However, using more modes does improve the reconstruction in the wake, and also removes the smaller-scale spurious structures to a large degree. This exercise serves as a verification that the method described here is able to accurately reconstruct the flow-field from a small number of DMD modes. It also demonstrates that the dynamics are governed by a few modes even in these complex vortex-dominated flows, which makes them good candidates for the development of lower-order models using such methods.

\subsubsection{Flow Analysis using DMD}
\label{sec:airfoil}
In the previous section, we established the ability of this method to correctly decompose the flow around an airfoil that is performing large-amplitude pitch oscillations into its DMD modes. We now analyze these DMD modes for airfoils performing a variety of pitch kinematics, with the aim of using the most relevant DMD modes to analyze the dominant flow structures and their associated timescales. As mentioned in section \ref{sec:intro}, the fluid-structure coupling can be studied in terms of the energy transfer between the flow and the airfoil. We showed that this in turn depends on the phase difference between the kinematics and the flow structures responsible for force/moment generation on the airfoil. Hence, the timing and phase of the relevant flow structures will be of primary interest in this discussion and will be exploited to gain insight into the flow-induced response from the decomposed flow.

\begin{table}[h!]
\centering
\begin{tabular}{ |c|c|c|r| } 
 \hline
 $  $& $\boldsymbol{f^*_p}$ & $\boldsymbol{\Delta \theta}$ & \multicolumn{1}{|c|}{$\boldsymbol{C_E}$} \\ [0.5ex]
 \hline \hline
 $(1)$ & $0.20$ & $04$ & $2.79 \times 10^{-3}$ \\ [0.5ex]
 $(2)$ & $0.20$ & $10$ & $3.57 \times 10^{-2}$ \\ [0.5ex]
 $(3)$ & $0.20$ & $25$ & $6.45 \times 10^{-2}$ \\ [0.5ex]
 $(4)$ & $0.20$ & $29$ & $8.16 \times 10^{-2}$ \\ [0.5ex]
 $(5)$ & $0.20$ & $40$ & $-8.94 \times 10^{-2}$ \\ [0.5ex]
 $(6)$ & $0.30$ & $29$ & $-2.48 \times 10^{-1}$ \\ [0.5ex]
 \hline
\end{tabular}
\caption{Table summarizing the oscillation frequency ($f^*_p$), amplitude ($\Delta \theta$), and energy transfer coefficient ($C_E$) for the pitching airfoil cases analyzed in this work. For all cases listed in the table $Re=1000$, the airfoil is pitching about $\theta_0=15^{\circ}$, and the instantaneous geometric angle of attack is given by $\theta = \theta_0 + \Delta \theta\sin(2\pi f^*_p t)$.}
\label{table:energy}
\end{table}

For an airfoil pitching sinusoidally at a single frequency it is easy to show that only the mode of the forcing at the oscillation frequency, and non-harmonics modes, contribute to the structural excitation. For the cases studied here, we have verified that the computed DMD spectrum does not contain non-harmonic modes of the oscillation frequency. Hence the only mode that contributes to the excitation is the mode at the oscillation frequency. For this reason, the ensuing discussion will focus on the DMD mode at the oscillation frequency, and particularly on the effect of varying the amplitude of oscillation at a fixed frequency. We choose this fixed frequency to be $f^*_p=0.20$ for most of the discussion, and amplitudes of oscillation from $\Delta \theta = 4^{\circ}$ to $\Delta \theta =40^{\circ}$ will be analyzed. We also discuss the effect of changing frequency of oscillation at fixed amplitude, to highlight the different timescales involved. The locations on the energy map of all cases discussed here were shown in Figure \ref{fig:energy_map}. For reference, Table \ref{table:energy} shows the kinematic parameters and energy transfer ($C_E$) calculated for each case.

\begin{figure}
    \centering
    \includegraphics[scale=1.1]{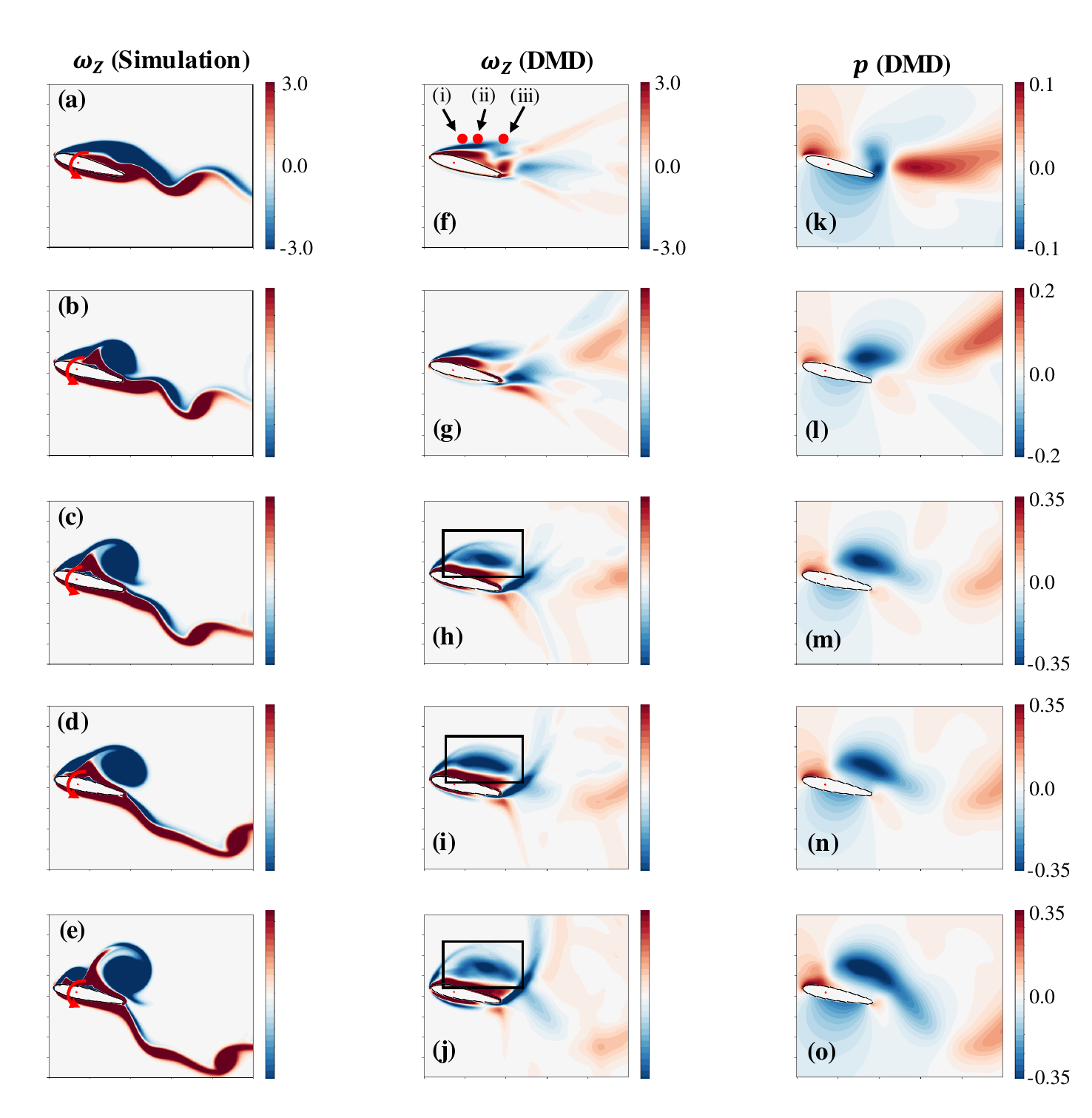}
    \caption{Analysis of selected cases from the energy map (cases shown in table \ref{table:energy}) using DMD modes of $Z$-vorticity and pressure. Cases shown correspond to various oscillation amplitudes at frequency of oscillation $f^*_p=0.20$ (cases (1)-(5) in table \ref{table:energy}). All snapshots are plotted at the time instance corresponding to the mean position during pitch-down. (a)-(e) $Z$-Vorticity contours from the simulation data for amplitudes of $\Delta \theta = 4^{\circ}$, $10^{\circ}$, $25^{\circ}$, $29^{\circ}$ and $40^{\circ}$ respectively; (f)-(j) DMD mode of vorticity at the oscillation frequency $(f^*=f^*_p)$, and amplitudes of oscillation correspond to those in (a)-(e). The rectangles in (h)-(j) highlight the flow structure of interest discussed in the text; (k)-(o) DMD mode of pressure at the oscillation frequency, and amplitudes of oscillation correspond to those in (a)-(e).}
    \label{fig:airfoil_20}
\end{figure}{}
We begin with a description of the full (undecomposed) vorticity field for the cases with $f^*_p=0.20$ and amplitude varying from $\Delta \theta=4^{\circ}$ to $\Delta \theta=40^{\circ}$. A snapshot of vorticity at the time-instance corresponding to the mean position during the pitch-down phase of the oscillation cycle for each of these cases is shown in figures \ref{fig:airfoil_20}(a)-(e). On increasing $\Delta \theta$ from $4^{\circ}$ to $40^{\circ}$ we see the growth, separation, and roll-up of the leading-edge shear layer as $\Delta \theta$ is increased. We also see the presence of strong vortex structures in the wake in all these cases. From observing these snapshots of the full flow field it is unclear what dominant flow structures correspond to the oscillation frequency in these cases. Further, we see from figures \ref{fig:airfoil_20}(c)-(e) that although the vortex structures in these cases have different sizes, they have qualitatively similar features. However these cases show different energy extraction, with the case of $\Delta \theta=40^{\circ}$, shown in figure \ref{fig:airfoil_20}(e), showing negative energy extraction, and all other cases showing positive energy extraction. Hence, from observing the full flow field it is not clear what causes the different energy transfer values. This motivates an analysis procedure based on decomposing the flow to extract the important feature that drive the dynamics.

In figures \ref{fig:airfoil_20}(f)-(j), we plot the DMD mode of $Z$-vorticity at the oscillation frequency for an airfoil pitching at $f^*_p=0.20$ and various amplitudes of oscillation. This is shown at the time-instance corresponding to the airfoil's mean position ($\theta=15^{\circ}$) during the pitch-down phase of the oscillation cycle. At small amplitude oscillations of $\Delta \theta = 4^\circ$, shown in figure \ref{fig:airfoil_20}(a), we see that the dominant vortex structures at the oscillation frequency are the separating shear layer near the leading edge, and prominent vortex structures near the trailing edge. This regime of small-amplitude oscillation shows vortex shedding that is very similar to that in bluff-bodies, which is evident from the trailing-edge shedding. On increasing the amplitude of oscillation to $\Delta \theta=10^{\circ}$, in figure \ref{fig:airfoil_20}(b), we see the development of a rolled-up shear layer over the suction surface of the airfoil, along with the initial development of an LEV at this amplitude of oscillation. The shear-layer and LEV at the oscillation frequency is presumably responsible for the large increase in energy transfer over the case with smaller amplitude oscillations, as seen in table \ref{table:energy}.

As we move to larger amplitude oscillations, from $\Delta \theta=25^{\circ}$ to $\Delta \theta=40^{\circ}$, we see that the dominant vortex structure is again the separating shear layer, which rolls up into an LEV-like structure over the suction surface. This structure is highlighted for $\Delta \theta=25^{\circ}$, $\Delta \theta=29^{\circ}$ and $\Delta \theta=40^{\circ}$ in figures \ref{fig:airfoil_20}(h)-(j) respectively, at the mean pitch-down position, which corresponds to the maximum angular velocity of the airfoil. It is particularly interesting to analyze the dominant vortex structure at this time-instance since, in the context of energy transfer, we expect its influence to be maximum (or close to maximum) when the angular velocity of the airfoil goes through a maximum. Hence it is instructive to compare this highlighted structure for the cases discussed here, i.e. $\Delta \theta=25^{\circ}$, $\Delta \theta=29^{\circ}$ and $\Delta \theta=40^{\circ}$, to understand how it affects the energy transfer as we increase the oscillation amplitude. 

In the case of $\Delta \theta=25^{\circ}$, shown in figure \ref{fig:airfoil_20}(h), we see that the strength of the LEV-like structure does indeed go through a maximum close to the time-instance shown (it must be noted that although we show just one snapshot for brevity, we have confirmed the occurrence of this maximum using time-resolved snapshots). On increasing the amplitude of oscillation to $\Delta \theta=29^{\circ}$, we again see the occurrence of a large LEV-like structure over the airfoil in figure \ref{fig:airfoil_20}(i), which goes through its maximum size as the angular velocity of the airfoil peaks. However, it is interesting to note that the extent of this structure is larger in the case of $\Delta \theta=29^{\circ}$ than for $\Delta \theta=25^{\circ}$, which is evident from comparing the highlighted structures in figures \ref{fig:airfoil_20}(h) and \ref{fig:airfoil_20}(i). This is in line with the fact that the energy transfer is larger at $\Delta \theta=29^{\circ}$, as seen in table \ref{fig:energy_map}(b). At $\Delta \theta=40^{\circ}$, which is shown in figure \ref{fig:airfoil_20}(j), we see similar structures as in the previous two cases. However, we again see that the spatial extent of the highlighted LEV-like structure is smaller than in the case of $\Delta \theta=29^{\circ}$ as the airfoil goes through its peak pitching velocity. Further, the vorticity over the suction-side of the airfoil extends farther downstream, which suggests that the LEV is shed earlier in the cycle at these larger amplitude oscillations.

Since we are interested in the fluid-structure coupling between the flow and the pitching airfoil, it is instructive to analyze the DMD modes of the pressure field in the flow, since pressure is the dominant forcing mechanism on the airfoil. Here we describe the DMD modes of the fluid pressure at $f^*=f^*_p=0.20$ for the cases discussed above, which corresponds to the oscillation frequency of the airfoil for these cases. In figure \ref{fig:airfoil_20}(k)-(o) we plot the DMD mode of the pressure field at the time-instance corresponding to maximum angular velocity of the airfoil during the downstroke. We see that for the small amplitude oscillation, $\Delta \theta=4^{\circ}$ in figure \ref{fig:airfoil_20}(k), there are alternating low and high-pressure structures shed off the suction and pressure sides of the trailing edge, similar to bluff-body shedding. One such trailing-edge vortex is evident at the time-instance plotted in figure \ref{fig:airfoil_20}(k). On increasing the amplitude of oscillation to $\Delta \theta=10^{\circ}$, shown in figure \ref{fig:airfoil_20}(l), there is evidence of a larger LEV over the suction-side of the airfoil. Further increasing the amplitude to $\Delta \theta=25^{\circ}$ leads to a larger low-pressure region over the airfoil, which we see in figure \ref{fig:airfoil_20}(m). The case of $\Delta \theta=29^{\circ}$ sees a still-larger low-pressure region over the airfoil, which is seen in figure \ref{fig:airfoil_20}(n). This is related to the occurrence of the larger shear-layer roll-up discussed previously for this case. At the highest amplitude of oscillation, $\Delta \theta=40^{\circ}$ shown in figure \ref{fig:airfoil_20}(o), we see that a large part of the low-pressure region over the airfoil has convected downstream of the airfoil, possibly influencing the pressure-side shear layer in an unfavourable manner at this time-instance. This supports our previous observation that the large-amplitude oscillation causes the shedding of vorticity earlier in the oscillating cycle, thus establishing an unfavourable phase difference and negative energy transfer. 

\begin{figure}
    \centering
    \includegraphics[scale=1.0]{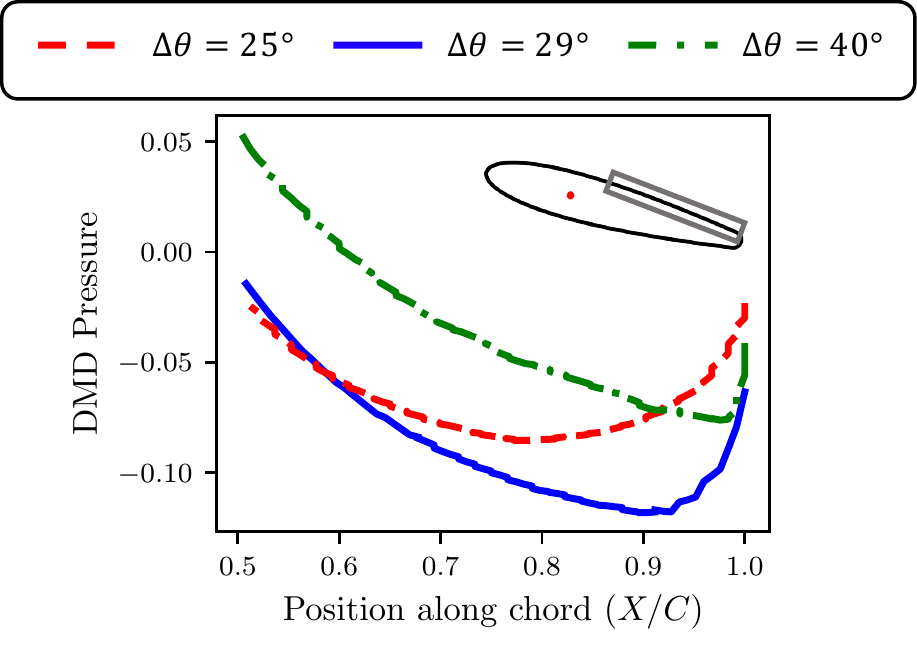}
    \caption{Surface pressure from the DMD mode at $f^*=f^*_p=0.20$, plotted along half the suction surface of the airfoil for three amplitudes of oscillation, $\Delta \theta = 25^{\circ}$, $\Delta \theta = 29^{\circ}$ and $\Delta \theta = 40^{\circ}$. The inset highlights the portion of the airfoil surface along which pressure is plotted. The pressure for the three cases is extracted from the pressure field shown in figures \ref{fig:airfoil_20}(m)-(o). }
    \label{fig:surface_pressure}
\end{figure}{}
In addition to the differences in the streamwise location of the pressure minimum discussed above, it is also interesting to note that there is a difference in the distance of this structure from the suction surface of the airfoil. We can see qualitatively from figures \ref{fig:airfoil_20}(m)-(o) that this pressure minimum is closest to the airfoil surface in the case of $\Delta \theta=29^{\circ}$ and farthest in the case of $\Delta \theta=40^{\circ}$. This can be verified by comparing the surface pressure induced by the DMD mode at $f^*=f^*_p=0.20$ for these three cases. In figure \ref{fig:surface_pressure} we plot this pressure along the airfoil surface, focusing on the portion of the suction surface closest to the pressure minimum (region of interest is highlighted in the inset of figure \ref{fig:surface_pressure}). We see that the mode at $f^*=f^*_p$ does indeed induce the lowest pressure on the surface in the case of $\Delta \theta=29^{\circ}$. Further, the pressure minimum for $\Delta \theta=40^{\circ}$ occurs farther downstream than in the other two cases, thus supporting our previous observation about the phase of this structure with respect to the motion. 

\begin{figure}
    \centering
    \includegraphics[scale=1.0]{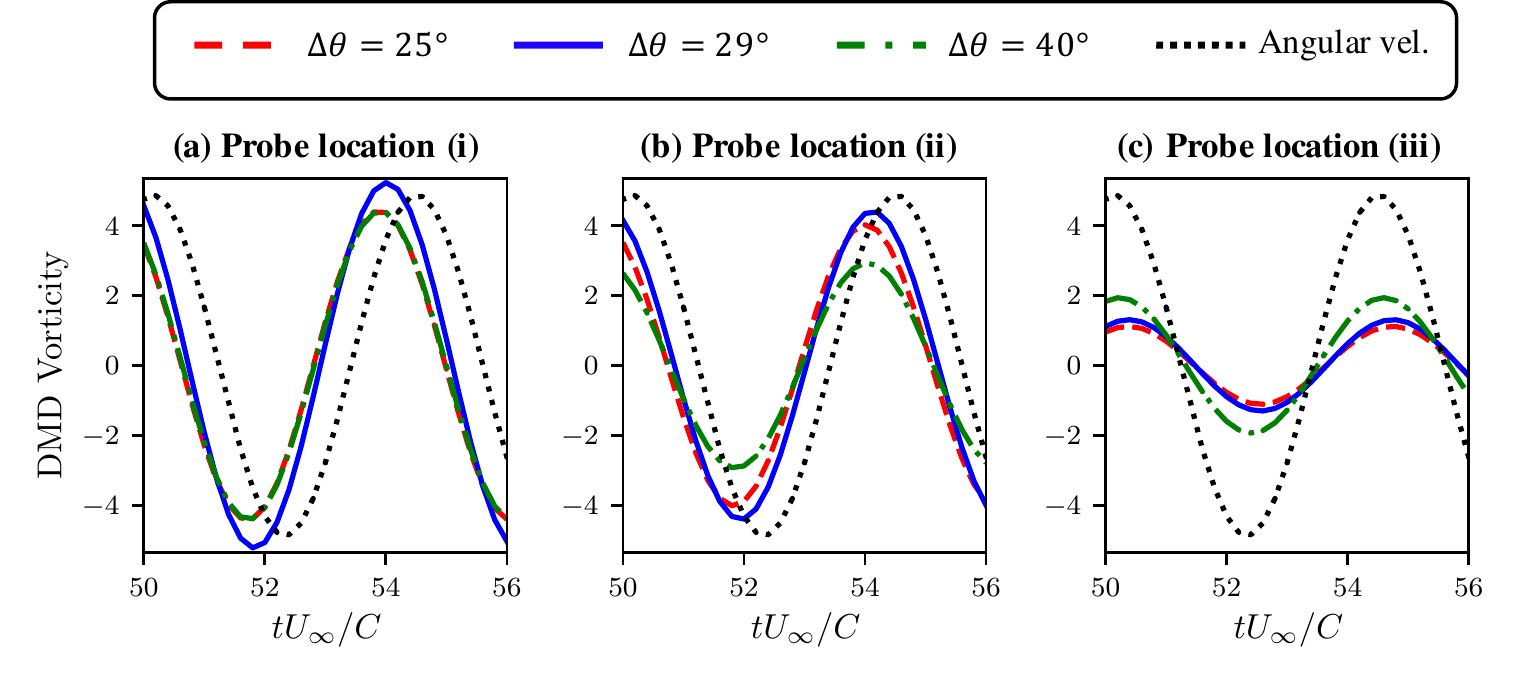}
    \caption{Time-series of $Z$-vorticity from the DMD mode corresponding to $f^*=f^*_p=0.20$, probed at three locations in the suction-side shear-layer. The probe locations (i),(ii) and (iii) are shown using red circles in figure \ref{fig:airfoil_20}(f) and the $Z$-vorticity is probed from the fields shown in figures \ref{fig:airfoil_20}(h)-(j). This is shown for cases with $\Delta \theta = 25^{\circ}$, $\Delta \theta = 29^{\circ}$ and $\Delta \theta = 40^{\circ}$. Also shown qualitatively (amplitude is arbitrary) using a dotted line is the angular velocity at $f^*_p=0.20$.}
    \label{fig:mode_probe}
\end{figure}{}
We can also make our argument about the phase and spatial extent of relevant vortex structures with respect to the motion of the airfoil more quantitative by probing individual points in the flow field over time. In figure \ref{fig:mode_probe}, we plot the value of vorticity from the DMD mode at $f^*=f^*_p=0.20$, i.e., the same DMD modes as in figures \ref{fig:airfoil_20}(f)-(j), at three points in the shear layer over time. Two positions at which we probe these values are located above mid-chord on the airfoil, separated by $0.1C$ in the streamwise direction, while the third point is slightly downstream of the trailing edge. These points are shown using red circles in figure \ref{fig:airfoil_20}(f). The time evolution of these quantities is compared with the angular velocity of the airfoil, which is shown qualitatively (focusing on phase with arbitrary amplitude) as a dotted line in the three plots. At probe location (i), corresponding to mid-chord, it is evident that the shear layer in the case of $\Delta \theta=29^{\circ}$ is both stronger (by about $20\%$) as well as slightly more in-phase with the angular velocity of the airfoil. Moving downstream to probe location (ii), the shear layer in the case of $\Delta \theta=29^{\circ}$ is still roughly $10\%$ stronger than that of $\Delta \theta=25^{\circ}$, and about $25\%$ stronger than that of $\Delta \theta=40^{\circ}$. However when we move downstream of the airfoil to location (iii), the shear layer in the case of $\Delta \theta=40^{\circ}$ is much larger than the other two cases, although it must be noted that the magnitude is significantly smaller at this downstream position. These measurements hence confirm the qualitative observations discussed previously, i.e., the strength (and to a lesser degree the phase) of the dominant vortex structure in the case of $\Delta \theta=29^{\circ}$ goes through a maximum more in phase with the maximum angular velocity of the pitching airfoil as compared to the other two cases discussed. Further, in the case of $\Delta \theta=40^{\circ}$ the dominant vortex structures are farther downstream as the airfoil goes through its maximum angular velocity. This is due to the fact that the large oscillation amplitude forces the generation and shedding of these structures earlier in the oscillation cycle.

We see from the above discussion that the DMD modes computed by this method over the pitching airfoil are able to provide valuable insight into the interaction between the vortex structures and airfoil motion. At the phase of maximum angular velocity of the airfoil, the qualitative difference in the phase of the separated shear layer in the three cases is able to highlight subtle differences that are not otherwise apparent when analyzing the full flow-field. Further, the decomposition of the pressure field also highlights important differences in the regions of minimum pressure and the resultant induced pressure on the airfoil surface. The maximum size and strength of the vortex structure and pressure minimum in the case of $\Delta \theta=29^{\circ}$ is in line with the fact that this case shows the maximum energy transfer. Further, the downstream position of the vortex convecting over the shear layer, as well as the relative position of the pressure minimum over the suction surface also suggest reasons for the the negative energy transfer in the case of $\Delta \theta=40^{\circ}$. 


\begin{figure}
    \centering
    \includegraphics[scale=1.0]{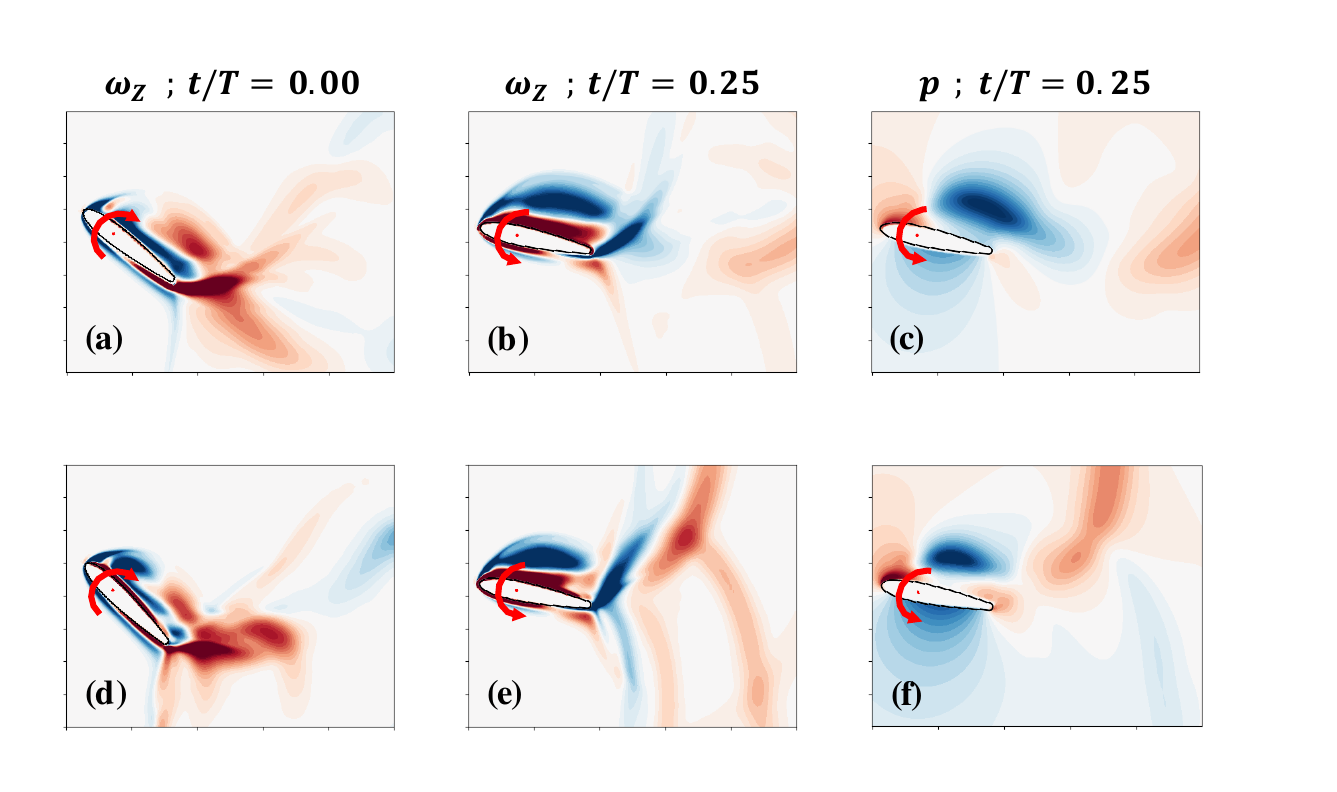}
    \caption{Comparison of DMD modes at the pitch frequency $(f^*=f^*_p)$ for airfoils oscillating with amplitude $\Delta \theta = 29^{\circ}$ and two different frequencies, $f^*_p=0.20$ and $f^*_p=0.30$. (a) $Z$-Vorticity contours of the DMD mode at the maximum pitch-up position for $f^*=f^*_p=0.20$; (b) $Z$-Vorticity contours of the DMD mode at the mean pitch-down position for $f^*=f^*_p=0.20$; (c) Pressure contours of the DMD mode at the mean pitch-down position for $f^*=f^*_p=0.20$; (d)-(f) Same quantities as in (a)-(c) but for $f^*=f^*_p=0.30$. Colour-bars for all $Z$-vorticity and pressure contours here are the same as in figure \ref{fig:airfoil_20}. }
    \label{fig:airfoil_20_30}
\end{figure}{}
We now analyze the effect of changing the frequency of oscillation while keeping the amplitude constant. Specifically, we discuss the DMD modes of the flow around airfoils oscillating with amplitude $\Delta \theta = 29^{\circ}$, and frequencies $f^*_p=0.20$ and $f^*_p=0.30$. In figure \ref{fig:airfoil_20_30} we show vorticity contours as well as pressure fields of the DMD mode corresponding to the oscillation frequency for each case. The top panel shows the case with  $f^*=f^*_p=0.20$ and the bottom panel shows the case with $f^*=f^*_p=0.30$. Figures \ref{fig:airfoil_20_30}(a) and \ref{fig:airfoil_20_30}(d) show the vorticity field at the time-instance corresponding to the peak pitch-up position. Comparing the two cases at this time instance, we see the generation of multiple vortices over a cycle in the case of $f^*=0.30$, which is particularly evident near the trailing edge and in the wake. Further, we also see evidence of this at the mean pitch down position, shown in figures \ref{fig:airfoil_20_30}(b) and \ref{fig:airfoil_20_30}(e), where the vortices convected into the wake show a fragmented structure. The pressure fields at the mean pitch-down position, shown in figures \ref{fig:airfoil_20_30}(c) and \ref{fig:airfoil_20_30}(f) also show a similar fragmented structure. In particular, we see that the low-pressure region over the suction side of the airfoil is much smaller in the case of $f^*=0.30$, and this is accompanied by a large high-pressure structure. The presence of these smaller-scale structures in the case of $f^*=0.30$ indicates that the timescale associated with the dominant flow-structures, the vortex shedding and shear-layer roll-up, is different from that of the oscillation. Comparing this to the fact that we observe increasingly smaller-scale structures at higher frequencies in figure \ref{fig:airfoil_reconst_modes}, the smaller-scale structures in the case of $f^*=0.30$ suggest that the dynamics of the flow occur at a lower fundamental frequency than the oscillation frequency of the airfoil. This discrepancy between timescales leads to negative energy transfer in this case. 

\section{Conclusions}
\label{sec:conclusions}
We have demonstrated a novel application of Dynamic Mode Decomposition (DMD) to fluid-structure interaction problems involving a rigid body of non-zero thickness performing large-amplitude motions. This formulation focuses on computing the DMD modes of the flow around the moving body in the lab-frame, where the motion of the body is taken into account. We have compared the results from this method and the standard DMD formulation for the flow around a periodically rotating circular cylinder, and showed under what conditions the results match. Specifically, we showed that the modes at frequencies not equal to the oscillation frequency are unmodified. For modes at the oscillation frequency, the method takes into consideration non-inertial effects in the velocity field, and retains some spatial effects of the moving reference frame. 
This method is applied to the analysis of the flow around an airfoil performing periodic pitching oscillations. We showed that this method is able to reconstruct the flow-field using a small number of modes (as is the case with the standard DMD), which is useful for the development of low-order models for these flow.  For the pitching airfoil, the modes at the oscillation frequency and the first harmonic contain the separating shear-layer, and its roll-up into LEV-like structures. The higher harmonics capture the convection of these vortices into the wake. Further, we were able to use the computed modes of the flow at the oscillation frequency to gain insight into the timing and phase of flow structures that interact with the airfoil during an oscillation cycle. We showed that by analyzing the mode corresponding to the oscillation frequency, at the time-instance when the angular velocity of the airfoil goes through a peak, we can gain significant insight into the phase relationship between the fluid and structure. We also showed the ability of this method to capture the difference in timescales between the oscillation frequency and flow structures, by analyzing the decomposition of the flow around airfoils pitching with the same amplitude but different frequencies. 

We expect that this method has applications in a variety of fluid-structure interaction problems that involve a body of non-zero thickness moving through a flow-field, particularly because it allows the analysis of the phase of the flow structures with respect to the moving body. However, the method does have drawbacks, one of which is the presence of spatial effects in the computed modes which originate from the decomposition in the moving frame. This is a possible direction for future work. 

\section*{Acknowledgments}
This work is supported by the Air Force Office of Scientific Research Grant Number FA 9550-16-1-0404, monitored by Dr. Gregg Abate. The development of the flow solver used here has benefited from NSF grants CBET-1511200 and PHY-1806689.



\section*{References}
\label{sec:references}
\bibliographystyle{elsarticle-harv} 
\bibliography{references}








\end{document}